\documentclass{article}
\usepackage{graphicx}
\begin{document}
\noindent

\title{CONFINEMENT OF COLOR: A REVIEW}
\bigskip
\author{Adriano Di Giacomo\\}
\maketitle
{\abstract{  The status of our understanding of the mechanisms of color
confinement is reviewed, in particular the results of numerical
simulations on the lattice.
}}

\section{Introduction}
Quarks and gluons are visible at short distances. They have never been
observed as free particles.
    Search for quarks started in 1963, when first Gell-Mann introduced them
as fundamental constituents of hadrons: the signature is their fractional
charge $q=\pm1/3,\pm2/3$.

    The upper limit to the ratio of quark abundance to proton abundance is
$n_q/n_p< 10^{-27}$\cite{PDG}, to be compared with the expectation in the
Standard Cosmological Model $n_q/n_p =  10^{-12 }$\cite{OK}.
    The experimental limit on the production cross section in nucleon
collisions is\cite{PDG}  $\sigma<10^{-40} {\rm cm}^2$ to be compared with the
expected value  $\sigma \sim 10^{-25} cm^2$ in the absence of confinement. The
only natural explanation of these small numbers is that these ratios are
exacly zero, or that confinement is an absolute property  due to some
symmetry.

    A transition, however, can occur at high temperature to a phase in
which quarks and gluons are deconfined and form a quark-gluon
plasma.\cite{CP}.

    Big experiments at heavy ions colliders aim to detect such a phase
transition, even if no clear signature for it is known.\cite{QM02}
    A number of theoretical ideas exist on the mechanism of confinement,
which we shall briefly review below.

    Lattice is a unique tool to investigate the existence of the deconfining
phase transition, to check the theoretical ideas and possibly to give
indications on what to look at in experiments.
\section{Lattice investigations}
The partition function of a field theory at temperature  $T$  is equal to
the euclidean Feynman integral extending in time from  0  to  $1/T$, with
periodic boundary conditions in time for bosons, antiperiodic for
fermions.
\begin{equation}
       Z\equiv Tr{ exp(-H/T) } =
\int[d\phi]\,\exp\left[-\int_0^{1/T}dt \int d^3x L(\vec x,t)\right]
\end{equation}

    Finite temperature QCD is simulated on a lattice $N_s^3\times N_t$ with
$N_s\gg N_t$. The temperature $T$ is given by
\begin{equation}
T = \frac{1}{N_t a(\beta)}
\end{equation}
where a is the lattice spacing in physical units , which by
renormalization group arguments in the weak coupling regime is given by
\begin{equation}
a(\beta) =\frac{1}{\Lambda} \exp(-\beta/b_0)
\end{equation}
   with  $\beta = 2N/g^2$; $-b_0$ the lowest order coefficient
of the beta function, which is negative because of asymptotic freedom. It
follows then from eq.(2) that the strong coupling region corresponds to
low temperatures, weak coupling to high temperatures.

    If a deconfining phase transition exists at some temperature, how can it
be detected, or what is the criterion for confinement?

    For pure gauge theories a reasonable answer exists, which consists in
looking at the static potential acting between a quark and an antiquark
at large distances:  if it is positive and diverging  there
is confinement by definition. In principle this criterion does not insure
that no colored particles exist as an asymptotic state, but  certainly
means that heavy quarks are confined.
    The static potential is related to the correlator $D(\vec x)$ of
Polyakov lines
\begin{equation}
    D(\vec x) =   \langle L(\vec x) L(0)\rangle
\end{equation}
     as follows
\begin{equation}
     V(x)= - T\ln D(\vec x)
\end{equation}
It can be shown by use of the cluster property that at large distances
\begin{equation}
    D(\vec x) \mathop\simeq_{|\vec x|\to\infty}
c\,\exp\left(-\frac{\sigma}{T}|\vec x|\right) + |\langle L\rangle|^2
\end{equation}
A temperature $T_c$ is found in numerical simulations such that
\begin{itemize}
\item[]
for $T < T_c$  $\langle L\rangle = 0$   and hence   $V(r) =  \sigma
\cdot r$ (confinement)
\item[]
for $T > T_c$  $\langle L\rangle\neq 0$,       $V(r) \sim const$
(deconfinement)
\end{itemize}
both for $SU(2)$  and for SU(3)  pure gauge theories.
$\langle L\rangle$ is an order parameter for confinement, and $Z_3$
is the corresponding
symmetry.

Of course no real phase transition can take place on a finite lattice
\cite{LY} , so that the transition from 0  to 1 of $\langle L\rangle$ is smooth
on a finite lattice , and becomes steeper and steeper as the size goes
large. The steepness is measured by the susceptibility $\chi_L $,
\begin{equation}
\chi_L = \int d^3 x\langle L(x) L^\dagger(0) -  L(0) L^\dagger(0)\rangle
\end{equation}
which diverges with some critical index $\gamma$ at the critical point
\begin{equation}
\chi_L\mathop\simeq_{\tau\to 0}\tau^{-\gamma}\qquad
\tau\equiv\left(1-\frac{T}{T_c}\right)
\end{equation}
Other relevant critical indices are the index $\nu$ of the correlation
length $\xi$ of the order parameter
\begin{equation}
\xi \propto \tau^{-\nu}
\end{equation}
and the index  $\alpha$ of the specific heat
\begin{equation}
   C_v - C^0_v \propto \tau^{-\alpha}
\end{equation}
$\alpha$,$\gamma$ and $\nu$ identify the universality class and/or the
order of the phase transition. A weak first order transition is a
limiting case $\alpha=1$ , $\gamma$ =1 and  $\nu$= 1/d (d the number of
spacial dimensions i.e. 3).

    The critical indices are determined from the dependence of
susceptibilities on the spacial size of the system, by use of a technique
known as finite size scaling\cite{F}.
    The result is that for quenched SU(2) the transition is second order
and belongs to the universality class of the 3d ising model\cite{SU2}, for
SU(3) it is weak 1rst order \cite{SU3}.

In the presence of dynamical quarks $Z_3$ is not a symmetry and therefore
the Polyakov line cannot be an order parameter.Moreover the string breaks
due to the instability for production of dynamical quark pairs and the
potential at large distances is not growing with the distance, even if
there is confinement.
Another symmetry exists at zero quark masses, the chiral symmetry.At T=0
it is spontaneously broken , the pseudoscalar bosons being the Goldstone
particles , but it is restored at $T\simeq 170$~Mev. The corresponding
order parameter is  $\langle\bar\psi\psi\rangle$.  It is not clear
what exacly chiral symmetry
has to do with confinement:  in any case it is explicitely broken by
quark masses , and therefore it cannot be the symmetry responsible for
confinement discussed in sect 1. For a theory with $N_f=2$, $m_u=m_d=m$,
which is a model  approximation of reality, the situation is
schematically represented
in fig 1. The critical  line $T_c(m)$ is defined by the maxima of the
susceptibilities  $\chi_L,\chi_{\bar\psi \psi}, \chi_{C_v}$, which
coincide within errors
\cite{KL,FU}, and as an empirical definition the region below the
line is assumed to be confined, the region above it to be deconfined.
Theoretical ideas are needed to understand the symmetry pattern  of the
system.

\begin{figure}
\includegraphics[scale=0.4]{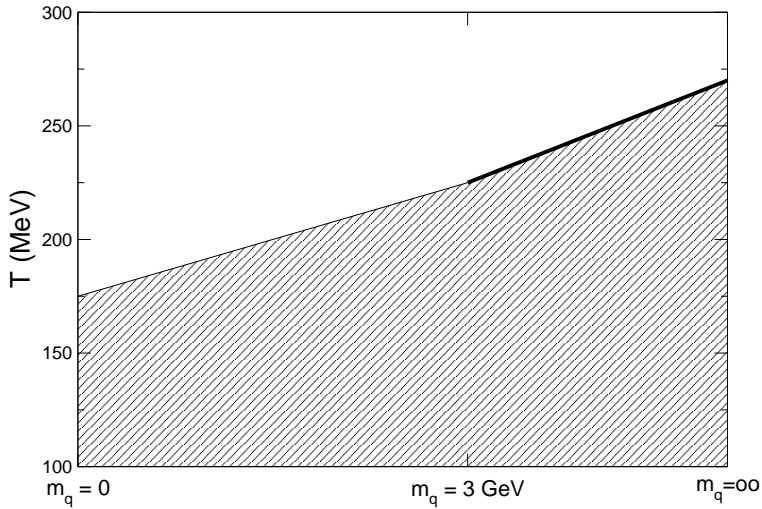}
\caption{The phase diagram of $N_f=2$ QCD.}
\end{figure}

As for the chiral transition a renormalization group analysis and the
assumption that that the pions are the relevant degrees of freedom gives
the following predictions\cite{PW}.  If the U(1) axial symmetry is
restored below the chiral transition, the transition is first order and
such is the critical line at $m\neq 0$. If instead the anomaly persists below
$T_c$ the transition is second order and the critical line at $m\neq0$ is a
crossover.

\section{Theoretical ideas}
A number of theoretical models of deconfinement exist in the literature.

There is a Gribov model, which is a clever picture of the chiral phase
transition\cite{Gr}. It does not apply, however, to quenched theory. In
the spirit of the $N_c\to\infty$ approach the mechanism of
confinement should be
the same for quenched and unquenched.

    Confinement could be produced by the condensation of vortices \cite{tH78}
The model corresponds to a well defined symmetry in 2+1 dimensions and
quenched theory , but  in any case the $Z_3$ symmetry does not survive
the  introduction of dynamical quarks.

    A most appealing idea is dual superconductivity of the vacuum\cite{tHM}:
chromoelectric charges are confined by dual Meissner effect, which
sqeezes the chromoelectric field acting between colored particles into
Abrikosov flux tubes, in the same way as magnetic charges are confined in
ordinary superconductors. A number of pioneering papers on this mechanism
were based on the definition and the counting of monopoles.\cite{ref}
    We shall instead concentrate on the symmetry patterns involved.
    Dual superconductivity means that the vacuum is a Bogolubov-Valatin
superposition of states with different monopole charge (monopole
condensation).

In order to define monopoles a magnetic U(1) gauge
symmetry must be identified in QCD , which has to be a color singlet if
monopoles condense without breaking the color symmetry.

The procedure to
identify such magnetic U(1) is known as Abelian Projection \cite{tH81}.
We shall present the abelian projection in a form which will prove useful
for what follows\cite{SUN}. Let  $G_{\mu\nu} = T^i G_{\mu\nu}^i$  be
the gauge field
strength  with $T^i$  the gauge group generators in the fundamental
representation, and  $\Phi=  T^i\Phi^i$ any operator in the adjoint
representation.
Define
\begin{equation}
F_{\mu\nu} =
Tr\left\{ \phi G_{\mu\nu}\right\} - \frac{i}{g}
Tr\left\{ \phi \left[D_\mu\phi, D_\nu\phi\right]\right\}
\end{equation}
$F_{\mu\nu}$  is gauge invariant and color singlet, and such are separately
the two terms in its definition.

    Theorem\cite{SUN}. A necessary and sufficient condition for the cancellation
of bilinear terms $A_\mu  A_\nu$ between the two terms in the right hand
side of eq.(11)  is that
\begin{equation}
         \Phi = \Phi^a =    U(x)\Phi^a_{diag} U^\dagger(x)
\end{equation}
with $U(x)$ an arbitrary gauge transformation and
\begin{equation}
\Phi_{diag}^a =
   diag\left(
\overbrace{\frac{N-a}{N},..,\frac{N-a}{N}}^{a},
\overbrace{-\frac{a}{N},..,-\frac{a}{N}}^{N-a}\right)\end{equation}

For any choice of the form eq(12)  $F_{\mu\nu}$ obeys Bianchi identities and
the identity holds
\begin{equation}
F^a_{\mu\nu} =
\partial_\mu Tr\left\{ \phi^a A_\nu\right\} -
\partial_\nu Tr\left\{ \phi^a A_\mu\right\}
-\frac{i}{g}
Tr\left\{ \phi^a \left[\partial_\mu\phi^a,
\partial_\nu\phi^a\right]\right\}\end{equation}

$F_{\mu\nu}$ is gauge invariant and can be computed in the gauge in
which $\Phi^a$
is diagonal. In that gauge

\begin{equation}
F^a_{\mu\nu} = \partial_\mu Tr(\Phi^a A_\nu) - \partial_\nu Tr(\Phi^a A_\mu)
\end{equation}
has an abelian form.
By developing  $A^\mu_{diag}$  in terms of roots  $A_\mu  =  \alpha^i A^i_\mu$
\begin{equation}
\alpha^i = diag(0,0,0\ldots\stackrel{i}{1},\stackrel{i+1}{-1},0\ldots 0)
\end{equation}
with
\[ Tr(\alpha^i \Phi^j) =\delta^{ij}\]
\begin{equation}
F^a_{\mu\nu} = \partial_\mu   A^a_\nu - \partial_\nu  A^a_\mu
\end{equation}
The gauge transformation $U(x)$ which brings
to the unitary gauge
is called abelian projection.

A magnetic current can be defined as
\begin{equation}
J_\nu^a =\partial_\mu F^{a *}_{\mu\nu}
\end {equation}

This current is identically zero due to Bianchi identities, but can be
non zero in compact formulations like lattice regularization , in which
Dirac strings are invisible. In any case  $J_\mu^a$   is identically
conserved
\begin{equation}
\partial_\mu J_\mu^a = 0
\end{equation}
and defines a magnetic U(1) conserved charge.
    This magnetic U(1) symmetry can either be Higgs broken, and then the
system is a magnetic ( dual) superconductor, or it can  be realized
\`a la Wigner,and then magnetic charge is superselected.
    For any choice of  the field $\Phi^a$ in eq (12 ) a magnetic U(1) 
symmetry is
defined.

    To detect dual superconductivity the vev of a magnetically charged
operator can be used as an order parameter. Such an operator has been
constructed \cite{refr},and is magnetically charged and U(1) gauge invariant
\cite{ref2} . The continuum version of the construction goes as follows.
Define
\begin{equation}
\mu^a(\vec x,t) =
e^{
i\int d^3\vec y\,Tr\left(\phi^a_{}\vec E(\vec y,t)
\right)
\vec b_\perp(\vec x -\vec y)
}
\end{equation}
where $\Phi^a$ is defined by eq(12) and $\vec E(\vec x,t)$ is the
chromoelectric field
operator  $E_i = G_{0i}$ and
\begin{equation}
\vec\nabla\vec b_\perp = 0\quad, \vec\nabla\wedge \vec b_\perp
= \frac{2\pi}{g}\frac{\vec r}{r^3} +\hbox{Dirac string}
\end{equation}
$\mu^a$ is gauge invariant by construction if $\Phi^a$
transforms in the adjoint representation. In the abelian projected gauge,
where $\Phi^a = \Phi^a_{diag}$   it assumes the form
\begin{equation}
\mu^a(\vec x,t) =
\exp\left\{i\int d^3\vec y\,{\vec E}_\perp^a(\vec y,t)
\vec b_\perp(\vec x -\vec y)\right\}
\end{equation}

where  ${\vec E}_\perp^a$ is the component of the electric field
along the residual $U(1)$ direction
as defined by eq.(16),(17),
and only the transverse part survives in the convolution with $\vec b_\perp$ .
    In any quantization procedure  ${\vec E}_\perp^a$ is the conjugate
momentum to  ${\vec A}_\perp^a$
so that $\mu^a$ is nothing but the translation operator of ${\vec
A}_\perp^a$ and
\begin{equation}
\mu^a(\vec x,t) | {\vec A}^a_\perp(\vec y,t)\rangle
=  | {\vec A}^a_\perp(\vec y,t) + \vec b_\perp(\vec x - \vec y)\rangle
\end{equation}
$\mu^a$ creates a magnetic monopole.

    In the confined phase , in which monopoles condense and the ground state
is not an eigenstate of the magnetic charge $\langle\mu_a\rangle\neq 0$
can signal dual superconductivity. In the deconfined phase
$\langle\mu_a\rangle = 0$
    All this refers to a given choice of the abelian projection, i.e. of
the gauge transfomation U(x) defining  $F^a_{\mu\nu}$ .

    To explore\cite{DG}\cite{DP} how physics depends on the choice of 
the abelian
projection let us go back to eq (20). By use of the cyclic invariance
of the trace
$\mu^a$ can be rewritten
\begin{equation}
\mu^a(\vec x,t) =
e^{
i\int d^3\vec y\,Tr\left(\phi^a_{diag}U^\dagger(\vec y,t)\vec E(\vec y,t)
U(\vec y,t)
\right)
\vec b_\perp(\vec x -\vec y)
}
\end{equation}
    In computing the correlation functions of $\mu^a$'s a change of variables
can be performed in the Feynman integral corresponding to a gauge
transformation generated by
U(x). If U(x) is independent of the field configuration the jacobian of
the transformation is  1  , the operator $\mu^a$assumes to all effects the
form  eq.(24)    so that the correlators , and in particular the one point
function $\langle\mu^a\rangle$ are independent of $U(x)$.
    If $U(x)$ depends on the field configuration , as happens e.g. for the max
abelian gauge or for any gauge in which a specific field dependent
operator is diagonalized, then the jacobian can be different from 1 and
the correlators depend on the abelian projection.
     However, if the number density of monopoles is finite ,the gauge
transformation which connects two abelian projections is continuous
everywhere except in a finite number of points and preserves topology: the
operator  $\mu^a$  defined by eq(22)  will then create a monopole in all
abelian projections. If $\langle\mu^a\rangle\neq0$  it signals dual
superconductivity in all abelian projections.

     An extensive investigation of the density of monopoles in different
abelian projections has been performed, and indeed the number density of
monopoles is finite. Fig 2 illustrates the method , and refers
specifically to the abelian projection in which the Polyakov line
operator is diagonal. The eigenvalues of that (unitary) operator have the
form
\begin{equation}
L_i = e^{i\phi_i}\qquad i=1,2,3
\end{equation}
and in defining the abelian projection are ordered in decreasing order of
$\phi_i$.  A monopole singularity in a point  x  implies that two eigenvalues
are equal , eg $\phi_1$ and $\phi_2$. Fig  2 shows the distribution of the
difference of the first two eigenvalues on the lattice sites of 1000
field configurations of quenched SU(3) on a $16^4$ lattice . In no site
there is a monopole. Repeating the determination on a finer lattice gives
similar results.
    As a consequence one can state that
the number density of monopoles is finite and the
dual superconductivity (or non) is
an intrinsic property , independent of the abelian projection which
defines the monopoles.

\begin{figure}
\includegraphics[angle=270,scale=0.4]{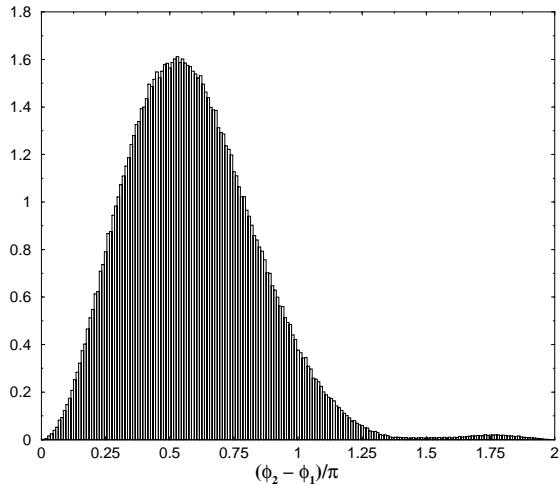}
\caption{An example of probability distribution of the difference of
the two highest
eigenvalues of the phase $\Phi$ of the  Polyakov line $e^{i\Phi}$, at
the lattice
sites. $SU(3)$ gauge group, $\beta = 6.4$, lattice $16^4$, $10^3$
configurations.}
\end{figure}

An extensive analysis on the lattice \cite{refr} shows that the vacuum is
indeed a dual superconductor in the confined phase, and goes to normal
in the deconfined one. A finite size scaling analysis of the
susceptibility  $\rho$  defined as
\begin{equation}
\rho^a = \frac{d}{d\beta}\ln\langle\mu^a\rangle
\end{equation}
gives  that in the quenched case  $\langle\mu^a\rangle$  is strictly zero
above the critical temperature defined by the Polyakov line order
parameter, is different from zero below it. The behaviour around $T_c$
allows to determine the critical indices , which are consistent with those
determined by use of the Polyakov line. This is clear evidence that
dual superconductivity is a mechanism for confinement\cite{refr}.

\section{The case of full QCD}
The order parameters $\langle\mu^a\rangle$ can equally well be
defined in the presence
of dynamical quarks  (Full QCD)  \cite{ref3} and have the same physical
meaning of creators of monopoles. One can then ask if a criterion for
confinement could be provided by $\langle\mu^a\rangle$'s , i.e. by
dual superconductivity
(or absence of) . One should prove that in the dual superconducting phase
no colored asymptotic states exist, which is of course non trivial.
However this is not much different from the situation in quenched theory
with the Polyakov line criterion , as discussed in sect 2.
    It has in fact been checked \cite{ref3} that QCD  vacuum is a dual
superconductor in the phase below the critical line of fig1,
($\langle\mu^a\rangle\neq0$),
and is normal in the region above it ($\langle\mu^a\rangle=0$).
    The finite size scaling analysis in this case  goes as follows.
For dimensional reasons the order parameter $\langle\mu^a\rangle$ has the
form
\begin{equation}
\langle\mu^a\rangle = \Phi(\frac{a}{\xi},\frac{N_s}{\xi},m L^{y_h})
\end{equation}
where $\xi$ is the correlation length, a the lattice spacing, m the
quark mass and $y_h$ the corresponding anomalous dimension .
    Near the critical line $\xi$ goes large compared to  a  and the
dependence on a/$\xi$ can be neglected. (Scaling)
    The problem has two scales. If $y_h$ is known one can choose different
values of the mass and of the spacial size $N_s$ such that m$N_s$ is
constant, and then
\begin{equation}
\langle\mu^a\rangle \mathop\simeq_{T\to T_c} f(\frac{N_s}{\xi})
\end{equation}
     By use of eq(9) the variable $N_s\xi$ can be traded with $\tau N_s^{1/\nu}$
and the scaling law follows
\begin{equation}
\rho^a = N_s^{1/\nu} f(\tau N_s^{1/\nu})\, , \qquad \tau = (1- \frac{T}{T_c})
\end{equation}
whence $\nu$ can be extracted and the order of the transition can be
determined. The result is $\nu = .33$ compatible with a first order transition.

A cross check is obtained by studying the scaling of the maximum of the
specific heat, which for the same choices of m and $N_s$ should scale as
\begin{equation}
C_v - C_v^0 \sim Ns^{\alpha/\nu}
\end{equation}
If the critical indices determined through $\langle\mu^a\rangle$
coincide with those
resulting from the analysis of the specific heat, this would be
additional evidence for dual superconductivity as a mechanism of
confinement, implying that $\langle\mu^a\rangle$ can be the order parameter.

The situation is described in ref\cite{PICA} ,and is presently at the stage
of indication that this is indeed the case. Numerical work is on the way
which will definitely clarify the problem.

    In conclusion Confinement is a fundamental but difficult problem.
Some understanding has been reached on the symmetry patterns involved.
Lattice is a unique tool to address the problem.

Thanks are due to my collaborators J.M. Carmona, L.Del Debbio, M.
D'Elia, B. Lucini, G. Paffuti, C. Pica for discussions .
This work is partially supported by MIUR Progetto Teoria delle
Interazioni Fondamentali.

\end{document}